\documentclass[%
 aip,
 amsmath,amssymb,
reprint,%
]{revtex4-1}

\usepackage{graphicx}
\usepackage{dcolumn}
\usepackage{lipsum}
\usepackage{bm}
\usepackage{tabu}
\usepackage[utf8]{inputenc}
\usepackage[T1]{fontenc}
\usepackage{mathptmx}
\usepackage{color}

\newcommand{\ket}[1]{\ensuremath{\left|#1\right\rangle}}
\newcommand{\braket}[2]{\ensuremath{\left\langle #1 | #2 \right\rangle}}

\begin{document}
\preprint{AIP/123-QED}
\title[Classically-entangled Ince-Gaussian modes]{Classically-entangled Ince-Gaussian modes}
\author{Yao-Li}
\author{Xiao-Bo Hu}
\affiliation{Wang Da-Heng Collaborative Innovation Center for Quantum manipulation \& Control, Harbin University of Science and Technology, Harbin 150080, China}%
\author{Benjamin Perez-Garcia}
\affiliation{Photonics and Mathematical Optics Group, Tecnologico de Monterrey, Monterrey 64849, Mexico}%
\author{Bo-Zhao}%
\author{Wei Gao}
\author{Zhi-Han Zhu}
\author{Carmelo Rosales-Guzm\'an}
\email{carmelorosalesg@hrbust.edu.cn}
\affiliation{Wang Da-Heng Collaborative Innovation Center for Quantum manipulation \& Control, Harbin University of Science and Technology, Harbin 150080, China}%

\date{\today} 

\begin{abstract}
Complex vector light modes, classically-entangled in their spatial and polarisation degrees of freedom (DoF), have become ubiquitous in a vast diversity of research fields. Crucially, while polarisation is limited to a bi-dimensional space, the spatial mode is unbounded, it can be specified by any of the sets of solutions the wave equation can support in the different coordinate systems. Here we report on a class of vector beams with elliptical symmetry where the spatial DoF is encoded in the Ince-Gaussian modes of the cylindrical elliptical coordinates. We outline their geometric representation on the Higher-Order Poincar\'e Sphere, demonstrate their experimental generation and analyse the quality of the generated modes via Stokes polarimetry. We anticipate that such vector modes will be of great relevance in applications, such as, optical manipulations, laser material processing and optical communications amongst others.  
\end{abstract}

\maketitle

Complex vector modes have gained popularity in recent time due to their unique properties, which have enable application in research fields, such as optical manipulations, high-resolution microscopy, optical metrology, classical and quantum communications, amongst many others \cite{Rosales2018Review,Roadmap,Ndagano2018,Toppel2014,Li2016,Ndagano2017,Milione2015e,Bhebhe2018,Hu2019,Wang2015}. One of their main features is the non-separability between the spatial and polarisation Degrees of Freedom (DoF), which gives rise to non-homogeneous transverse polarisation distributions \cite{Galvez2003,Galvez2012}. Crucially, while the polarisation DoF is limited to a two-dimensional space, the spatial DoF is unbounded, any set of solutions of the wave equation can be used as an encoding basis. Common examples of such spatial modes are, Bessel- and Laguerre-Gaussian, Mathieu- and Ince-Gaussian (IG) or Hermite-Gaussian modes, of the cylindrical, elliptic cylindrical and Cartesian coordinate systems, respectively\cite{Siegman,Bandres2004,Bandres2008}. While vector modes with cylindrical symmetry have been widely studied, \cite{Zhan2009,Niv2004,Dudley2013,Otte2018b,Ren2015}, vector modes with elliptical symmetry have been only considered recently, in the context of singularity networks \cite{Otte2018a}. IG modes are natural solutions of the paraxial wave equation when solved in the elliptical cylindrical coordinates ${\bf r}=(\xi,\eta,z)$, where, $\xi\in[0,\infty)$ and $\eta\in[0,2\pi)$ are the radial and angular elliptical coordinates. The transverse coordinates are related to the Cartesian system $(x,y)$ according to $x = \sqrt{\frac{2}{\varepsilon \omega(z)^2}}\cosh\xi\cosh\eta$ and $y = \sqrt{\frac{2}{\varepsilon \omega(z)^2}}\sinh\xi\sinh\eta$, where $\omega(z)$ is the beam radius as function of the propagation distance $z$. Such modes can be classified according to their parity into even and odd and are described mathematically in terms of the even and odd Ince polynomials, $C_p^m(\cdot,\varepsilon)$ and $S_p^m(\cdot,\varepsilon)$, respectively, as \cite{Bandres2004}
\begin{equation}\label{IGscalar}
\begin{split}
    IG_{p,m,\varepsilon}^e({\bf r}) &= \frac{C\omega_0}{\omega(z)}C_p^m(i\xi,\varepsilon)C_p^m(\eta,\varepsilon)\text{e}^{-\frac{r^2}{\omega(z)}}\text{e}^{-i\left(kz+Z-\Phi\right)},\\
    IG_{p,m,\varepsilon}^o({\bf r}) &= \frac{S\omega_0}{\omega(z)}S_p^m(i\xi,\varepsilon)S_p^m(\eta,\varepsilon)\text{e}^{-\frac{r^2}{\omega(z)}}\text{e}^{-i\left(kz+Z-\Phi\right)},\\
    IG_{p,m,\varepsilon}^h({\bf r}) &= IG_{p,m,\varepsilon}^e({\bf r}) + i IG_{p,m,\varepsilon}^o({\bf r}),
\end{split}
\end{equation}
where $C$ and $S$ are normalisation constants. The superscripts $e$, $o$ and $h$ refer to the even, odd and helical modes, respectively. The indexes $p,m\in\mathbb{N}$ obey the elations $0\leq m\leq p$ for even functions and $1\leq m\leq p$ for odd functions. Further, $z_R=\pi\omega_0^2/\lambda$ is the Rayleigh length, $\Phi=(p+1)\arctan(z/z_R)$ is the Gouy phase and $Z(z)=kr^2/2R(z)$ is an additional phase term related to the radius of curvature $R(z)=z+z_R^2/z$ of the phase front. A characteristic factor of IG modes is their ellipticity, $\varepsilon=2f_0/\omega_0^2$, with $\omega_0$ the Gaussian beamwaist, which can vary in the interval $[0,\infty)$. This parameter allows a smooth transition between the Laguerre-Gaussian ($LG_q^\ell$) and Hermite Gaussian ($HG_{n_x n_y}$) modes. More precisely, for $\varepsilon=0$ the $IG_{p,m,\varepsilon}^{h}$ becomes the $LG_q^\ell$, with their indices related by $\ell=m$ and $q=(p-m)/2$. When $\varepsilon\to\infty$, the $IG_{p,m,\varepsilon}^{e,o}$ transform into the $HG_{n_x n_y}$, through the relations $n_x=m$, $n_y=p-m$,  for even modes and $n_x=m-1$, $n_y=p-m+1$ for odd modes \cite{Bandres2004b}.

In this manuscript we introduce the set of IG vector modes, clasically-entangled in their spatial and polarisation DoF, which are generated as a weighted superposition of the even and odd IG modes with orthogonal circular polarisation. This superposition allows a geometric representation of such modes on a Higher-Order Poincare Sphere (HOPS). Such geometric representations are very useful and therefore, they have been used previously to represent scalar IG modes, at both, the classical and quantum level \cite{Bandres2004b,Kren2013,Shen2019}. Further, we demonstrate the experimental generation of IG vector modes using digital holography implemented through a Digital Micromirror Device (DMD) \cite{Gong2014,Mitchell2016}. We also reconstruct their transverse polarisation distribution, via Stokes polarimetry and compare this with theoretical predictions, to assess their quality. We anticipate that the IG vector modes introduced here, featuring a variety of elliptical shapes and polarisation distributions will be of utmost importance in advanced applications.

In general, vector modes are generated as a weighted superposition of the spatial and polarisation DoF \cite{Rosales2018Review}. Here, the spatial DoF is encoded in the infinite set of modes defined by Eq. \ref{IGscalar}. Given that IG modes form a complete orthogonal basis, we can use any two modes to perform such superposition. Here, we will only analyse the particular case where the spatial DoF is encoded in the $IG_{p,m,\varepsilon_1}^e({\bf r})$ and $IG_{p,m,\varepsilon_2}^o({\bf r})$ mode sets. This case is of particular interest since it allows a natural transition between both sets via a weighted superposition of them that gives rise to the IG vector modes. The most general case in which the spatial DoF is encoded in the helical IG modes deserves its own analysis, which is an undergoing research and will be reported separately. It suffices to say that the case $m=p=1$ leads to the well--known cylindrical vector vortex modes, whereas higher order helical IG modes produce more exotic polarisation distributions.
\begin{figure}[t]
    \centering
    \includegraphics[width=.45\textwidth]{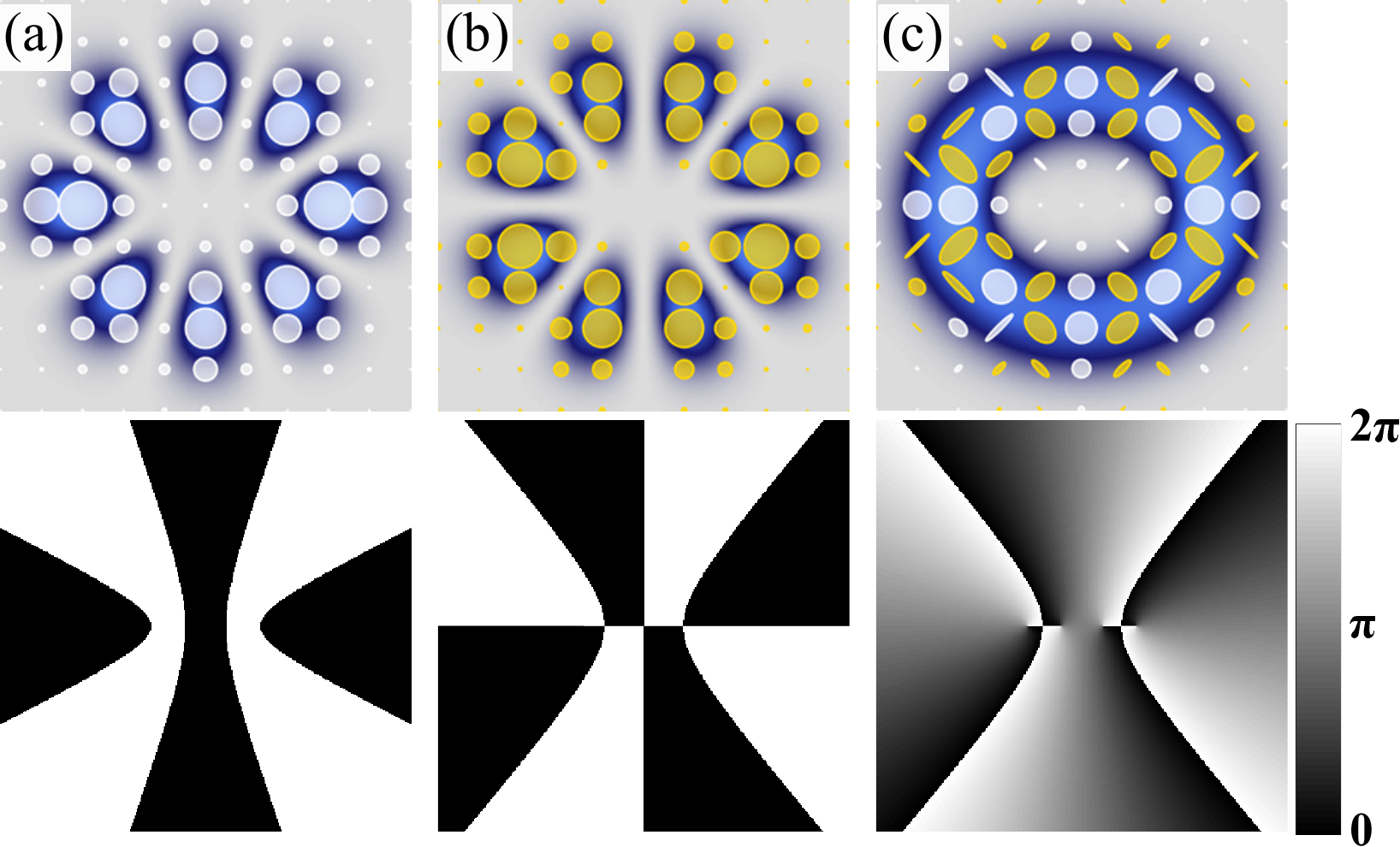}
    \caption{Transverse intensity distribution of (a) even mode with right (b) and odd mode with left circular polarisation. (c) IG vector modes showing a non-homogeneous transverse polarisation distribution.}
    \label{IGvector}
\end{figure}
The similarities between classical and quantum entangled states allow us to express the paraxial IG modes using Dirac's notation as\cite{Goyal2013a},
\begin{align}
\nonumber
    \braket{\xi,\eta}{p,m}_{\varepsilon}^e &= IG_{p,m,\varepsilon}^e({\bf r}),\\
    \braket{\xi,\eta}{p,m}_{\varepsilon}^o &= IG_{p,m,\varepsilon}^o({\bf r}),\\\nonumber
    \braket{\xi,\eta}{p,m}_{\varepsilon}^h &= IG_{p,m,\varepsilon}^h({\bf r}),
\end{align}
obeying the orthogonality rules $^{\sigma'}_\varepsilon\hspace{-4pt}\braket{p',m'}{p,m}_\varepsilon^{\sigma} = \delta_{\sigma,\sigma'}\delta_{p,p'}\delta_{m,m'}$, for $\sigma=\{e,o\}$. This in turn, allows the IG vector modes to be written as,
\begin{align}\label{InceModes}
    \ket{\Psi_{p,m}}_{\varepsilon} = \cos\theta\ket{p,m}_{\varepsilon}^e\ket{R} + \sin\theta\text{e}^{i\alpha}\ket{p,m}_{\varepsilon}^o\ket{L}.
\end{align}
Here, the kets $|R\rangle$ and $|L\rangle$ represent the right and left circular polarisation states while the kets $\ket{p,m}_{\varepsilon}^e$, $\ket{p,m}_{\varepsilon}^o$ the even and odd IG modes. The parameter $\theta \in [0, \pi/2]$ is a weighting factor that allows the field $|\Psi_{p,m}\rangle_\varepsilon$ to monotonically change from the even ($\theta=0$) to the odd ($\theta=\pi/2$) IG modes, passing through pure IG vector modes at $\theta=\pi/4$, as schematically represented in Fig. \ref{IGvector}, for the specific case $\ket{\Psi_{4,4}}_1$. Here, the intensity profile overlapped with the corresponding polarisation distribution is shown on the top row, while their phase profile on the bottom row. Figures \ref{IGvector}(a) and \ref{IGvector}(b) represent the scalar modes $\ket{4,4}_{1}^e\ket{R}$ and $\ket{4,4}_{1}^o\ket{L}$, respectively, while Fig. \ref{IGvector}(c) the vector mode  $\ket{\Psi_{4,4}}_{1}$ for the case $\alpha=0$. Notice the non-homogeneous polarisation distribution as well as the azimuthal phase variation. Additional parameters in Eq. \ref{InceModes} are the intramodal phase ${\text e}^{i\alpha}$ ($\alpha\in[0,\pi]$). In principle, this superposition can be done for any values of $p$, $m$ or $\varepsilon$, but suitable combinations produce propagation-invariant modes, characterised by the same Gouy phase \cite{Otte2018a}, which are more desirable in certain applications. The ellipticicy $\varepsilon$ is an additional parameter that might give rise to interesting properties, such as, the generation of hybrid vector modes resulting from the superposition of LG and HG modes.

\begin{figure}[b]
    \centering
    \includegraphics[width=.49\textwidth]{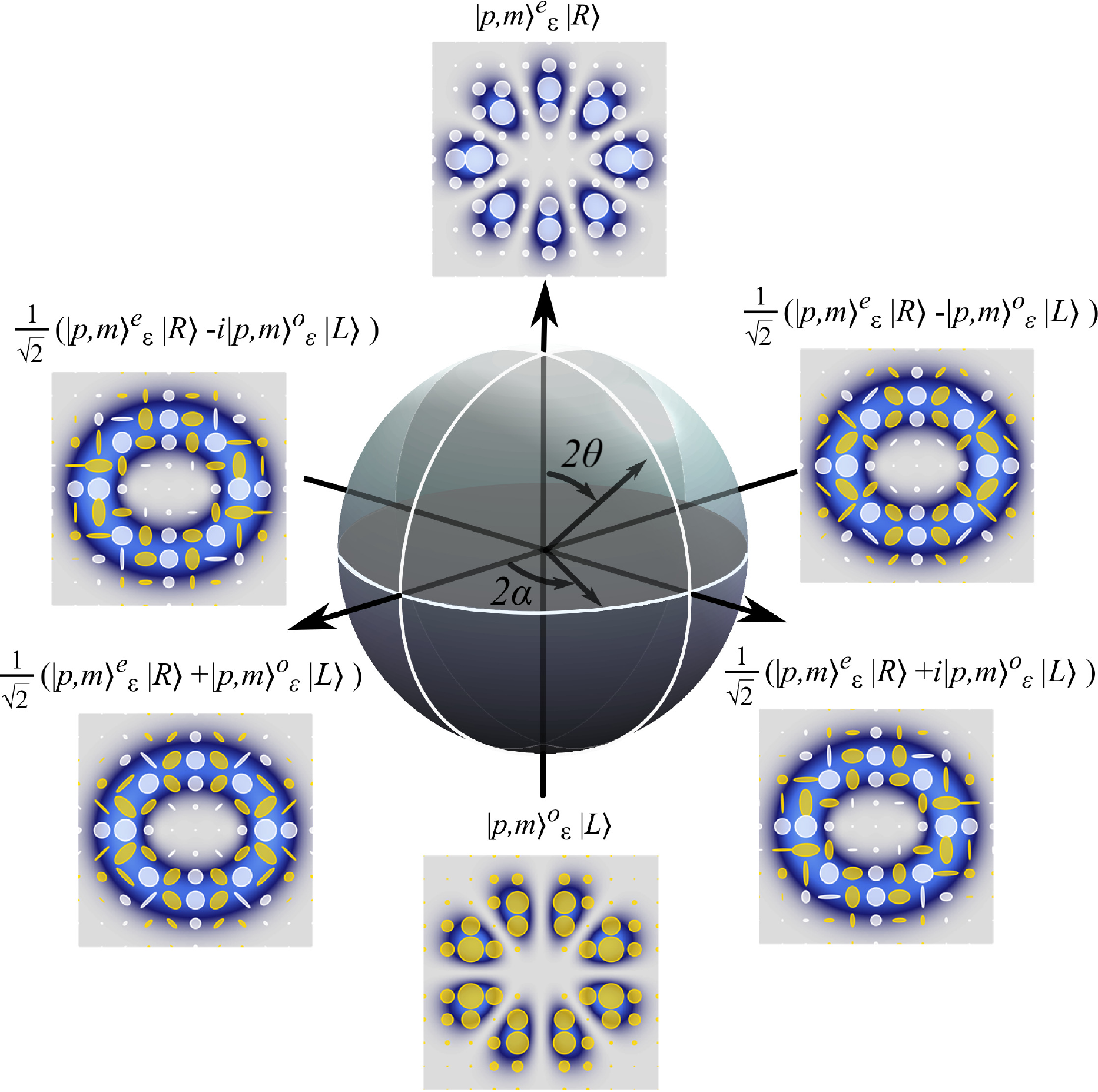}
    \caption{Geometric representation of IG vector modes on the HOPS. Even modes with right and odd modes with left states of circular polarisation are located in the North and South pole, respectively, while pure vector modes lie along the equator.}
    \label{HOPS}
\end{figure}

The vector states given by Eq. \ref{InceModes} can be represented geometrically on the well-known HOPS, wherein each vector mode is assigned to a unique point ($2\alpha$,$2\theta$) on the surface of a unitary sphere which assigns the North and South poles to the scalar modes $IG_{p,m,\epsilon}^{e}({\bf r})$ and $IG_{p,m,\epsilon}^{o}({\bf r})$, respectively. Pure IG vector modes are represented along the equator, whereas the remaining points feature intermediate states with non--equal weightings. Figure \ref{HOPS} illustrates this geometric representation for the specific case $\ket{\Psi_{4,4}}_1$, where the intensity profile overlapped with polarisation distribution is shown for representative cases along the equator and the poles.
\begin{figure*}[tb]
    \centering
    \includegraphics[width=.8\textwidth]{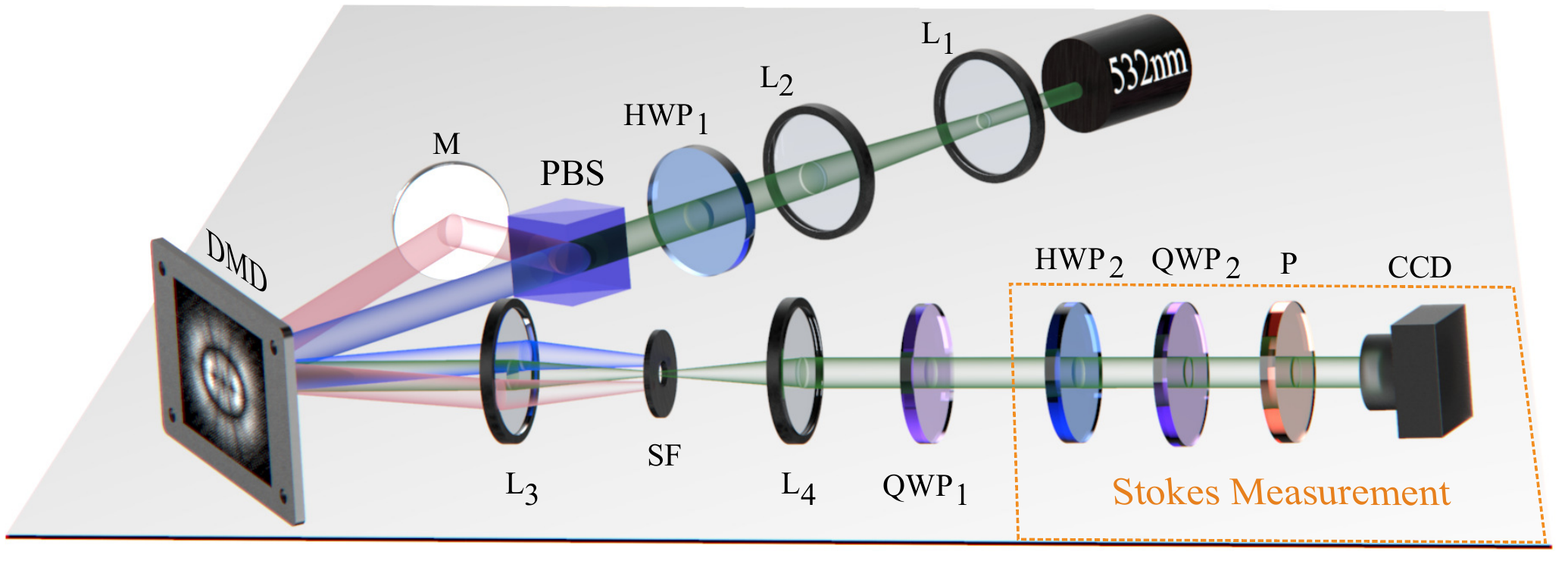}
    \caption{Experimental setup for the generation of IG vector modes. A diagonally polarised laser beam ($\lambda=532$nm), polarised with a Half-Wave plate (HWP$_1$) at 45$^{\circ}$, is expanded and collimated with lenses L$_1$ and L$_2$. It is then split by a Polarising beam splitter (PBS) into its vertical and horizontal polarisation components, which are redirected to the centre of a Digital Micro Device (DMD). Here two independent holograms with unique spatial frequencies are multiplexed into a single one, to generate the vector beam in the first diffraction order. Lenses L$_3$ and L$_4$ in combination with a Spatial Filter (SF), remove undesired diffraction orders. A Quater-Wave plate (QWP) transforms the polarisation basis from linear to circular. The polarisation distribution is reconstructed via Stokes polarimetry using a Charged-Coupled Device (CCD) camera.}
    \label{setup}
\end{figure*}

\begin{figure}[tb]
    \centering
    \includegraphics[width=0.47\textwidth]{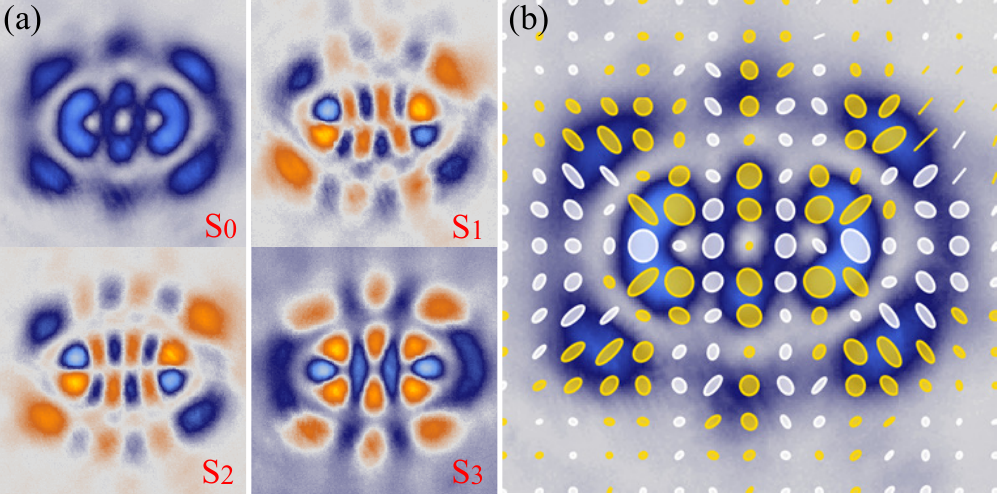}
    \caption{(a) Experimental Stokes parameters $S_0$, $S_1$, $S_2$ and $S_3$ of the mode $\ket{\Psi_{5,3}}_{2}$. (b) Reconstructed polarisation distribution.}
    \label{Fig:stokes}
\end{figure}

The experimental generation of IG vector modes was achieved through a compact experimental setup based on a DMD \cite{Selyem2019,Rosales2020}, schematically illustrated in Fig. \ref{setup}. Here, a continuous wave (CW) laser beam ($\lambda=532$ nm) with diagonal polarisation, achieved with a Half-wave plate (HWP$_1$) at $45^\circ$, was expanded and collimated using L1 $(f=50 \text{ mm})$ and L2 $(f=400 \text{ mm})$, to approximate a flat wavefront. A polarising beam splitter separates the beam into its two polarisation components, horizontal and vertical. The resulting beams impinge onto a polarisation-independent DMD (DLP Light Crafter 6500 from Texas Instruments) at two slightly different angles (approximately separated by $1.5^\circ$). Here, a binary hologram consisting of two superimposed (multiplexing) independent holograms with unique spatial carrier frequencies is displayed \cite{Rosales2017}, so that, one hologram generates the even while the other the odd mode. The spatial frequency of each hologram is selected to ensure the overlap of the first diffraction order of each beam along a common propagation axis, which is spatially filtered using the telescope formed by the lenses L3 and L4 ($f_{3,4}=200$ mm) in combination with a spatial filter (SP). This telescope images the DMD plane to the plane $z=0$, where the IG vector beams are analysed. A Quarter-wave plate (QWP$_1$) oriented at $45^o$ is added along the path of beam to transform it into the circular polarisation basis. The transverse polarisation distribution was reconstructed using Stokes polarimetry according to \cite{Goldstein2011}
\begin{equation}\label{Eq:Stokes}
\begin{split}
\centering
     &S_{0}=I_{0},\hspace{19mm} S_{1}=2I_{H}-S_{0},\hspace{1mm}\\
     &S_{2}=2I_{D}-S_{0},\hspace{10mm} S_{3}=2I_{R}-S_{0},
\end{split}
\end{equation}
where $I_0$ is the total intensity of the mode and $I_H$, $I_D$ and $I_R$ the intensity of the horizontal, diagonal and right-handed polarisation components, respectively. The intensities were acquired through the combination of a linear polariser (P), and a Quarter-wave plate (QWP$_2$) and recorded with a CCD camera, as detailed in\cite{Zhao2019}. $I_H$ and $I_D$ were measured by passing the IG vector field $|\Psi_{p,m}\rangle$ through a linear polariser at $0^\circ$ and $45^\circ$, respectively. $I_R$ was measured by passing the beam through a QWP at $45^\circ$ and a linear polariser at $90^\circ$. Figure \ref{Fig:stokes} shows an example of such measurements for the specific case $\ket{\Psi_{5,3}}_2$. Here, the Stokes parameters $S_{0}$, $S_{1}$, $S_{2}$ and $S_{3}$ are shown in Fig. \ref{Fig:stokes}(a), whereas the corresponding intensity profile overlapped with the polarisation distribution in Fig. \ref{Fig:stokes} (b).

\begin{figure}[tb]
    \centering
    \includegraphics[width=.47\textwidth]{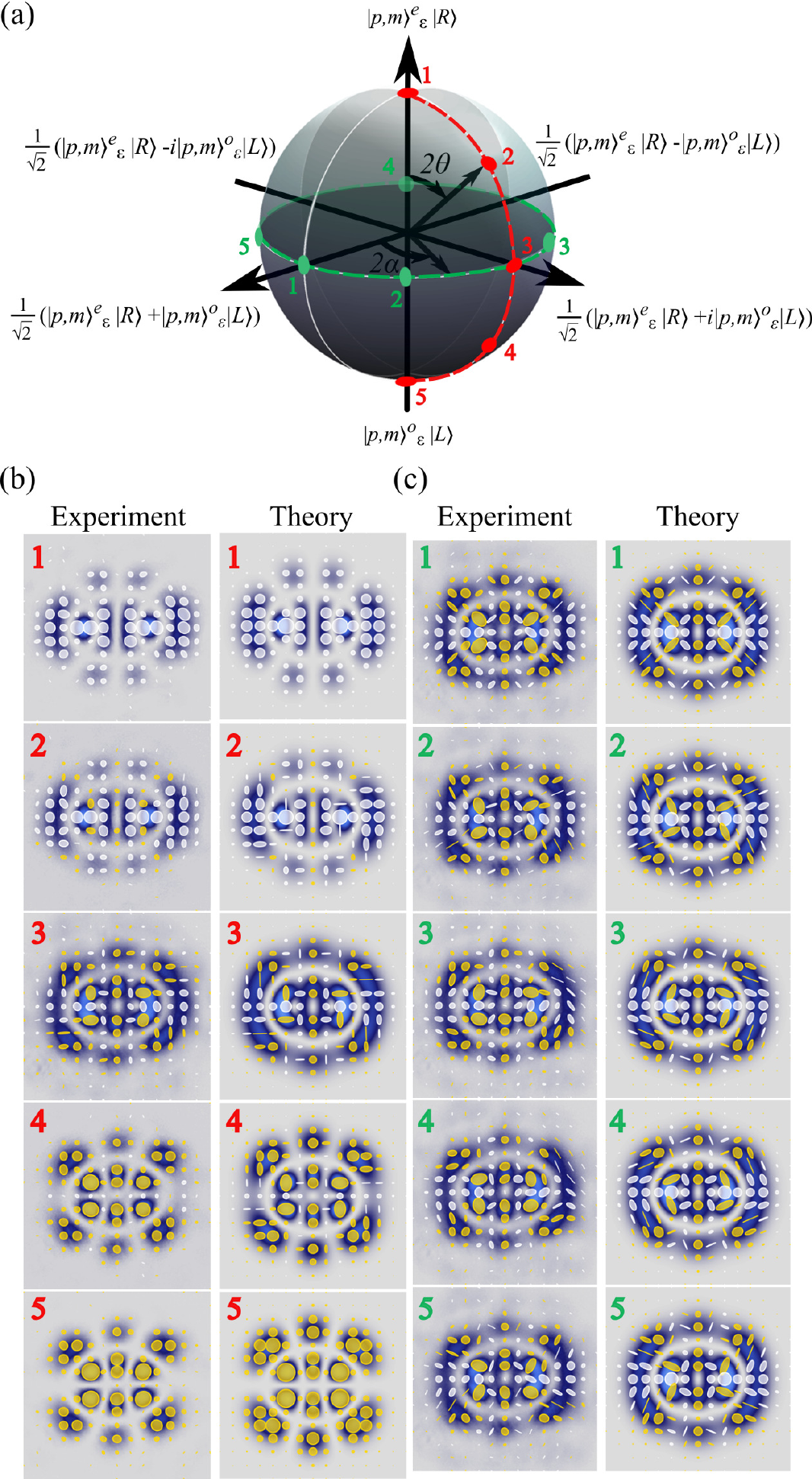}
    \caption{(a) Geometric representation of IG vector modes on the HOPS. Intensity and polarisation distribution of modes with, (b) varying degrees of non-separability and (c) different intramodal phases for the specific case $\ket{\Psi_{5,3}}_{2}$.}
    \label{Poincare1}
\end{figure}

With the setup described above we generated arbitrary IG vector modes, which are analysed in this section. First, we show the generation of arbitrary modes on the HOPS, using as away of example the set of IG vector modes given by $\ket{\Psi_{5,3}}_{3}$. The case of modes with different degrees of non-separability, represented on the HOPS along a path connecting the North and South poles, are schematically represented by the red dashed line shown in Fig. \ref{Poincare1} (a). Only a set of five representative modes are shown here, labelled with the numbers [1,2,3,4,5], which correspond to $2\theta\in[0,\pi/4,\pi/2,3\pi/4,\pi]$ and $\alpha=\pi/4$. Their transverse intensity profile, overlapped with their corresponding polarisation distribution are shown in the left panel of Fig.~\ref{Poincare1} (b). For comparison, their theoretical counterpart is shown on the left panel of the same figure. Notice the transition of these modes, from scalar (Fig. \ref{Poincare1}(b)-1) to vector (Fig.~\ref{Poincare1}(b)-3) and then back to scalar (Fig. \ref{Poincare1}(b)-5). 
As a second example, we show the set of pure vector modes generated with different intramodal phases, which are represented along the equator of the HOPS. Again, only a representative set of five modes are shown here, labelled with numbers from one to five. To experimentally generate such modes, we vary the intramodal phase $\alpha\in[0, \pi/8, 3\pi/8, 5\pi/8, 7\pi/8]$ while keeping $\theta$ constant. The experimental transverse intensity profile, overlapped with the corresponding polarisation distribution of such modes is shown in Fig.~\ref{Poincare1}(c), experiment on the left panel and theory on the right. Notice how the polarisation distribution rotates clockwise as $\alpha$ increases.

To quantify the deviation of the modes generated experimentally with theoretical predictions, we compared their transverse polarisation distribution across the entire transverse plane. For this, we used the root mean squared error (RMSE) of the orientation ($\alpha$) and flattening ($f$) of each polarisation ellipse (see \cite{Perez-Garcia2017} for more details). The left panel of table \ref{RMSE1}, shows the values corresponding to Fig. \ref{Poincare1}(b), while the right panel those correspond to Fig.\ref{Poincare1}(c). Notice that in all cases the $RMSE_\alpha$ is smaller than 7\%, while the $RMSE_f$ is smaller than 2\%, which evinces the accuracy of our generation method. 
\begin{table}[tb] 
\centering
\setlength{\tabcolsep}{4pt}
\renewcommand*{\arraystretch}{1.5}
 \caption{RMSE of IG vector modes on the HOPS.
 \label{RMSE1}}
 \begin{tabular}{c| c c c }
$2\theta$ & $RMSE_{\alpha}$ & $RMSE_f$ & \\
\hline
0 & 4.48\% & 0.79\% &\\ 
$\pi/4$ & 6.11\%& 0.83\% &\\
$\pi/2$ & 5.13\% & 1.50\% &\\
$3\pi/4$ & 5.53\% & 0.85\% &\\ 
$\pi$ & 3.97\% & 0.62\% &\\
\end{tabular}
\qquad
 \begin{tabular}{c| c c c }
$2\alpha$ & $RMSE_{\alpha}$ & $RMSE_f$ & \\
\hline
0 & 5.13\% & 1.26\% &\\ 
$\pi/4$ & 6.46\% & 1.32\% & \\
$3\pi/4$ & 6.57\% & 1.50\% & \\
$5\pi/4$ & 6.21\% & 1.32\% &\\ 
$7\pi/4$ & 5.71\% & 1.33\% & \\
\end{tabular}
\end{table}

\begin{figure}[h]
    \centering
    \includegraphics[width=0.47\textwidth]{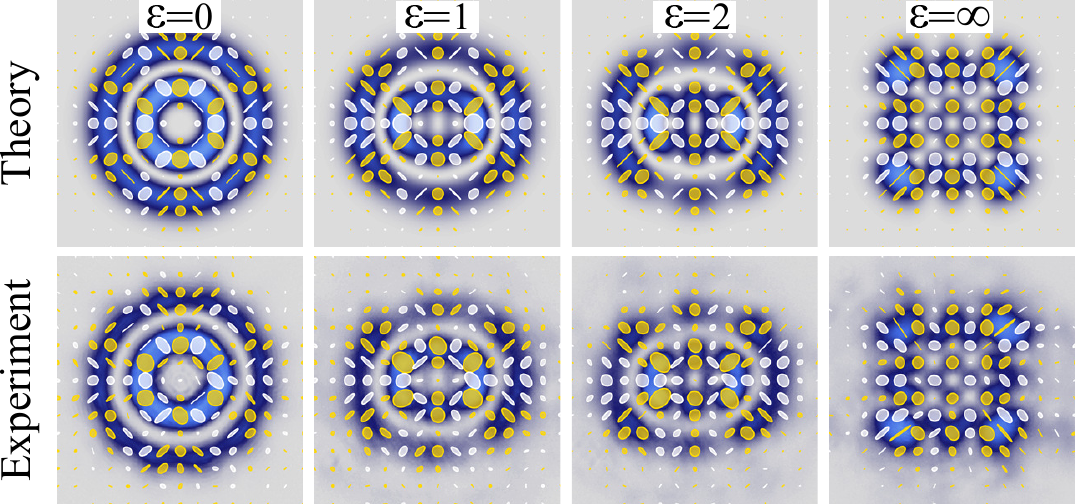}
    \caption{Polarization distribution of IG vector modes as function of the ellipticity $\varepsilon$. Theoretical (top) and experimental (bottom) transverse intensity profile overlapped with the polarisation distribution. Here we show the specific case $\ket{\Psi_{5,3}}_{\varepsilon}$ for $\varepsilon\in[0,1,2,\infty]$.}
    \label{Eccentricity}
\end{figure}

As a final example, we analysed the case of IG vector modes with increasing values of ellipticity $\varepsilon$. A representative example of the IG vector modes as function of $\varepsilon$ is shown in Fig. \ref{Eccentricity} for the specific cases $\ket{\Psi_{5,3}}_\varepsilon$ with, $\epsilon\in[0,1,2,\infty]$. Figure \ref{Eccentricity} shows a comparison of the transverse intensity profile overlapped with the corresponding polarisation distribution, theory in the top row and experiment in the bottom row. Again, we compared the transverse polarisation distribution of IG vector modes generated experimentally with their theoretical counterpart by computing the RMSE. The values obtained are displayed in table \ref{RMSE2} for each of the cases shown in Fig. \ref{Eccentricity}. 

\begin{table}[h] 
\centering
\setlength{\tabcolsep}{7pt}
\renewcommand*{\arraystretch}{2}
 \caption{RMSE of IG vector modes with different ellipticity.
 \label{RMSE2}}
 \begin{tabular}{c c c c c c}
Vector mode & $\varepsilon=0$ & $\varepsilon=1$ & $\varepsilon=2$ & $\varepsilon=\infty$ & \\
\hline
$RMSE_{\alpha}$ & 5.94\% & 5.60\% & 5.13\% & 5.50\% &\\ 
$RMSE_f$ & 1.29\% & 1.13\% & 1.26\% & 0.89\% &\\
\end{tabular}
\end{table}
In summary, here we have introduced the Ince-Gaussian (IG) vector modes as the superposition of the spatial DoF, encoded in the even and odd IG modes, and the polarisation DoF encoded in the circular polarisation basis. More precisely, we proposed a weighted superposition of even modes carrying right circular polarisation with odd modes carrying left circular polarisation. In this way, by varying the weighting coefficients we can achieve a monotonic transition from the even to the odd modes with left and right circular polarisation, respectively. Such superposition can be represented geometrically on the well-known Higher-Order Poincar\'e Sphere. In this representation, each IG vector mode is associated to a unique point on the surface of a unitary sphere, wherein the scalar even and odd modes are located in the North and South pole, respectively, wherein as the equator only pure vector modes are located. We generated such modes experimentally using a compact setup based on a polarisation-insensitive DMD for a fast and cheap generation. Further, we characterised the quality of the generated modes by reconstructing their transverse polarisation distribution using Stokes Polarimetry. To assess the quality of our generated modes, we compared their polarisation distribution across the transverse plane with theoretical predictions. To this end, we computed the RMSE of the orientation angle and flatness of each polarization ellipse. We anticipate that IG vector modes will pave the way to applications in fields such as optical metrology, optical manipulations and optical communications, to mention a few.

\section*{AUTHOR'S CONTRIBUTIONS}
Y.-L. and X.-B.H. contributed equally to this work.
\section*{Data availability}
The data that support the findings of this study are available from the corresponding author upon reasonable request.

This work was partially supported by the National Natural Science Foundation of China (NSFC) under Grant Nos.  61975047, 11934013, 11574065.

\section*{References}
%

\end{document}